%% file: main.tex
\documentclass[fleqn,usenatbib]{mnras}

\usepackage{newtxtext,newtxmath}

\usepackage[T1]{fontenc}

\DeclareRobustCommand{\VAN}[3]{#2}
\let\VANthebibliography\thebibliography
\def\thebibliography{\DeclareRobustCommand{\VAN}[3]{##3}\VANthebibliography}

\usepackage{graphicx}	
\usepackage{amsmath}	

\title[Efficiency of the oxygenic photosynthesis]{Efficiency of the oxygenic photosynthesis on Earth-like planets in the habitable zone}

\author[G. Covone et al.]{
Giovanni Covone,$^{1,2,3}$\thanks{E-mail: giovanni.covone@unina.it}
Riccardo M. Ienco,$^{1}$
Luca Cacciapuoti$^{1}$
and Laura Inno$^{2,4}$
\\
$^{1}$Dipartimento di Fisica "E. Pancini", Università di Napoli Federico II, Via Cinthia I-80126 Napoli, Italy\\
$^{2}$INAF, Osservatorio Astronomico di Capodimonte, Salita Moiariello, Napoli, Italy\\
$^{3}$INFN, Sezione di Napoli, C.U. Monte S. Angelo,
Via Cinthia, I-80126 Napoli, Italy\\
$^{4}$Science and Technology Department, Parthenope University of Naples, Naples, Italy
}

\date{Accepted XXX. Received YYY; in original form ZZZ}

\pubyear{2021}

\begin{document}
\label{firstpage}
\pagerange{\pageref{firstpage}--\pageref{lastpage}}
\maketitle

\begin{abstract}
Oxygenic photosynthesis is the most important biochemical process in Earth biosphere and likely very common on other habitable terrestrial planets, given the general availability of its input chemical ingredients and of light as source of energy. It is therefore important to evaluate the effective possibility of oxygenic photosynthesis on planets around stars as a function of their spectral type and the planet-star separation.
We aim at estimating the photon flux, the exergy and the exergetic efficiency of the radiation in the wavelength range useful for the oxygenic photosynthesis as a function of the host star effective temperature and planet-star separation.
We compute analytically these quantities and compare our results with the estimates for the small sample of known Earth-like planets.
We find that exergy is an increasing function of the star effective temperature, within the range 2600-7200 K. It depends both on the star-planet separation and the star effective temperature. 
Biospheres on exoplanets around cool stars might be generally light-limited. So far, we have not observed terrestrial planets comparable to Earth in terms of useful photon flux, exergy and exergetic efficiency.
\end{abstract}




\section{Introduction}

Search for life in the Universe is one of the most important and challenging scientific endeavours of our time. 
In particular, search for complex life and intelligence beyond the Solar System appears to be within reach by means
of the next generation of astronomical facilities
which will allow to detect biosignatures and technosignatures on nearby exoplanets, see e.g. \cite{Kiang_2018AsBio, Wright2019BAAS}.

For this reason, much research work has been focused on the study of the physical conditions favourable to complex life,
under the assumption that it is based on 
the availability of elements such as carbon, hydrogen, oxygen and nitrogen and water as a solvent system.
Indeed, 
although life based on non-aqueous solvents (as ammonia or methane) at low temperatures is considered to be possible, see e.g. \cite{Benner_2004},
water is believed to be an essential element for complex life. 
Therefore, the so called habitable zone \citep{dole1964habitable} for exoplanets
has been defined as the range of distances from the parent star
where a planet with a suitable atmosphere can host liquid water on its surface. 
Several authors have proposed climate models to describe the exoplanet atmospheres as a function of parent star luminosity,
star-planet distance and chemical composition of the atmosphere
in order to predict accurately the 
range of distances of the circumstellar habitable zone
(HZ, hereafter).

However, complex life requires much more stringent conditions on stellar and planet properties.   
For instance, \cite{Schwieterman_2019ApJ...878...19S} have recently shown that the HZ for complex aerobic life is likely more limited than previously thought because of too high 
CO$_2$ concentrations
at the HZ external edges (where the CO$_2$ greenhouse effect is required) or because of the large presence of toxic gases such as carbon monoxide.

The study of the physical conditions which might favour 
oxygenic photosynthesis (OP, hereafter) could also 
strongly constrain the parameter space for complex life.
Indeed, complex life on Earth is fully dependent on OP:
on Earth most autotrophic organisms synthesise organic matter from inorganic elements by using the energy from solar radiation (photosynthesis), only a limited group of organism use chemical energy (chemosynthesis), see e.g. \cite{Nakamura2015}.
Photosynthesis is the dominant process, as it allows to produce about 99\% of the entire biomass of the Earth biosphere, see e.g., \cite{jorgensen2004towards}.
OP is also essential for providing
abundant O$_2$ levels which appear to be necessary for the high-energy demands of multicellular life anywhere in the Universe,   
see \cite{Catling2005, McKay2014}. 

OP is a complex biochemical mechanism, but the overall reaction looks simple: %
\begin{equation}
6 \, \mathrm{CO}_{2}
+ 6 \, \mathrm{H}_{2} \mathrm{O} 
+ {\rm light}
\,  \rightarrow \, 
 \mathrm{C}_{6} \mathrm{H}_{12} \mathrm{O}_{6}+
6 \, \mathrm{O}_{2}
\label{eq:OP}
\end{equation}
We conjecture that
the chemical reaction (\ref{eq:OP})
should be quite common in the cosmos because of the generally large amounts of radiation received by exoplanets from their host stars, the  
availability of the input ingredients and its overall simplicity.
This view is also supported by the fact that OP
evolved very early on Earth.
The origin of OP is still highly debated, but it can be traced to a common progenitor of 
present-day cyanobacteria
earlier than about 2.4 billion years ago, 
the epoch of the so-called Great Oxidation Event, 
(see, e.g., \cite{Tomitani_2006})
or possibly even 3.6 billion years ago 
to an ancestral lineage of bacteria
\citep{Anbar2007, Cardona2018} (see also 
\cite{Fischer2016AREPS} for a recent review on this problem).

For this reason, several authors have studied the possibility for OP to take place in different extraterrestrial environments. 
\cite{Wald_1959} was  among the firsts to consider the generic properties of OP on other worlds.
\cite{Rueda1973} applied the second law of thermodynamics to derive bounds to the efficiency of exobiological photosynthesis.
More recent works on this topic include
\cite{Kiang_2007AsBio...7..252K},
\cite{Kiang_2007_AsBio...7..222K},
\cite{Lingam2019a},
\cite{Lehmer2018} and \cite{Claudi-life11010010}.
\cite{Ritchie2018I} presented a detailed analysis of the both oxygenic and anoxygenic photosynthesis on Proxima Centauri b.

In this work we study how the OP on other worlds would depend on
the stellar radiation from host stars on the main sequence, with effective temperature between 2600 and 7200 K.
In particular, we focus
both on the quantity of radiation (i.e., the photon flux) and its quality 
on Earth-like planets in the HZ.
In order to do this we use the notion of exergy,
a thermodynamic function that allows to quantify the maximum obtainable work from radiation.
Exergetic analysis is widespread in engineering and ecology whenever a cost analysis is required,
and it is been recently introduced by \cite{Scharf2019}
in the analysis of useful work from stellar radiation.
Similar studies on the OP on exoplanets
have been recently presented by  
\cite{Lingam2019a} and \cite{Scharf2019}. 
Our work differs in some important details of the models:
we do not limit our analysis to planets at the distance from the star 
where they could receive the same flux the Earth receives from the Sun, as we consider Earth-like planets in the entire habitable zone .

The outline of the paper is as follows. 
In Sect. 2 we compute the photon flux at the surface of  Earth-analogs in the HZ as a function of the effective host-star temperature
and we compare this with the required photon flux for OP as measured on Earth. 
In Sect. 3 we discuss the exergy flux and 
the exergetic efficiency of the stellar radiation at the planet position as a function 
of the host-star temperature. 
Finally, in Sect. 4 we summarize our conclusions.

\section{Photon flux for the oxygenic photosynthesis}

We focus on the OP on Earth-like planets in the HZ around main sequence stars with effective temperature between 2600 and 7200 K. 
This range is determined by the HZ model by \cite{Kopparapu2013} (see below) and 
includes the stars
with long enough lifetime to allow evolution of complex life.

The circumstellar HZ is traditionally defined as the region around the host star where a rocky planet with an Earth-like atmosphere can sustain liquid water on its surface, see, e.g., \cite{kasting1993habitable}.
In particular, we adopt the HZ model presented by \cite{Kopparapu2013} and \cite{Kopparapu2014}. 
In short, 
the authors assume a 1D, radiative–convective, cloud-free climate model. 
The HZ inner and outer edges are determined by  
atmospheres that are ${\rm H}_2 {\rm O}$ and ${\rm CO}_2$ dominated, respectively, 
with N$_2$ as a background gas. 
Therefore, the HZ inner and outer limits are defined by the so-called runaway greenhouse and maximum greenhouse effects.
We refer the reader to 
\cite{Kopparapu2013} for details of the climate models.

\cite{Kopparapu2014} studied the dependence of the HZ borders  
on the planet mass, 
considering the mass range 
between  0.1 and 5 Earth masses.
They find that in general, more massive planets have a wider HZ
as the atmospheric column depth scales with planet mass.
The main effect is for the inner edge, where more massive planets 
could receive about 10\% lower than Earth flux, thus allowing a more extended HZ towards the star.
In this work, we neglect this effect (i.e., we consider a model  with planets of about 1 Earth mass), as this does not impact our conclusions.

In the following we use the empirical relation 
proposed by \cite{Kopparapu2013}
between the host-star luminosity $L$ (and hence, its effective temperature) to the distances defining the 
inner and outer edges of the HZ :
\begin{equation}
d=\left(\frac{L / L_{\odot}}{S_{\mathrm{eff}}}\right)^{1/2} \mathrm{AU} \, ,
\label{eq:koppa}    
\end{equation}
where the adimensional parameter $S_{\rm eff} (T)$
is defined as the ratio between the net outgoing IR flux and the net incident stellar flux (both
evaluated at the top of the atmosphere). 
Note that, in this way, we implicitly assume that the planet orbit is fully confined in the HZ.

We assume the host star to emit blackbody radiation, with effective temperature $T$. 


Most photosynthetic organisms on Earth use the radiation in the wavelength range between about 400 and 700 nm \citep{field1998primary}. 
The so-called photosynthetically active radiation (PAR)  
is determined both by physical constraints and by evolutionary adaption, hence we need to understand how universal the PAR wavelength range is.
The short wavelength limit is determined by the high energy of UV photons that have ionizing effect and can damage the vegetable cells, see e.g.  \cite{Carvalho2011_UV}.
Therefore this limit is generally assumed to be universal.

It is well-known that some cyanobacteria, 
as {\it Acaryochloris marina}, 
use Chlorophyll d as primary photosynthetic pigment to harvest at wavelengths larger than 700 nm for OP \citep{Miyashita1996Natur}
and are wide-spread in Earth natural habitats  \citep{Zhang2019_Chl_d}.
More recently, \cite{Chen2010Science} discovered another photosynthetic pigment (Chlorophyll f) which allows organisms to further extend into the infrared region 
light-harvesting for OP.
%
%
Therefore, several authors  \citep[e.g.,][]{Ritchie2018I} have assumed the PAR to extend up to 750 nm.
Even such a small extension in the red would be significant
for worlds around cool stars.
\cite{Ritchie2018I} have found
that the overall irradiance in the range 400-700 nm is about 
63 $\mu$mol photon $\, {\rm m}^{-2} \, {\rm s}^{-1}$,
while in the range 700–749 nm
(potentially used by Chl d-type oxygenic organisms)
is about a factor of two larger, 
about 132 $\mu$mol photon $\, {\rm m}^{-2} \, {\rm s}^{-1}$.
However there is growing evidence of Earth organisms able to
use solar radiation up to about 
800 nm, 
see \cite{PETTAI2005, Kurashov2019}.
As on planets around cool stars, OP might be likely 
light-limited (see below),
these results suggest that PAR could be extended up to
$\sim 800$ nm, as a consequence of the strong evolutionary drive to develop the capacity to harvest long-wavelength radiation.
Therefore, we adopt 800 nm as the upper limit for the PAR in the following,
and compare our results with those obtained assuming the upper limit to be 750 nm as 
assumed in \cite{Ritchie2018I}.

\begin{figure*}
\centering
\includegraphics[width=0.85\textwidth]{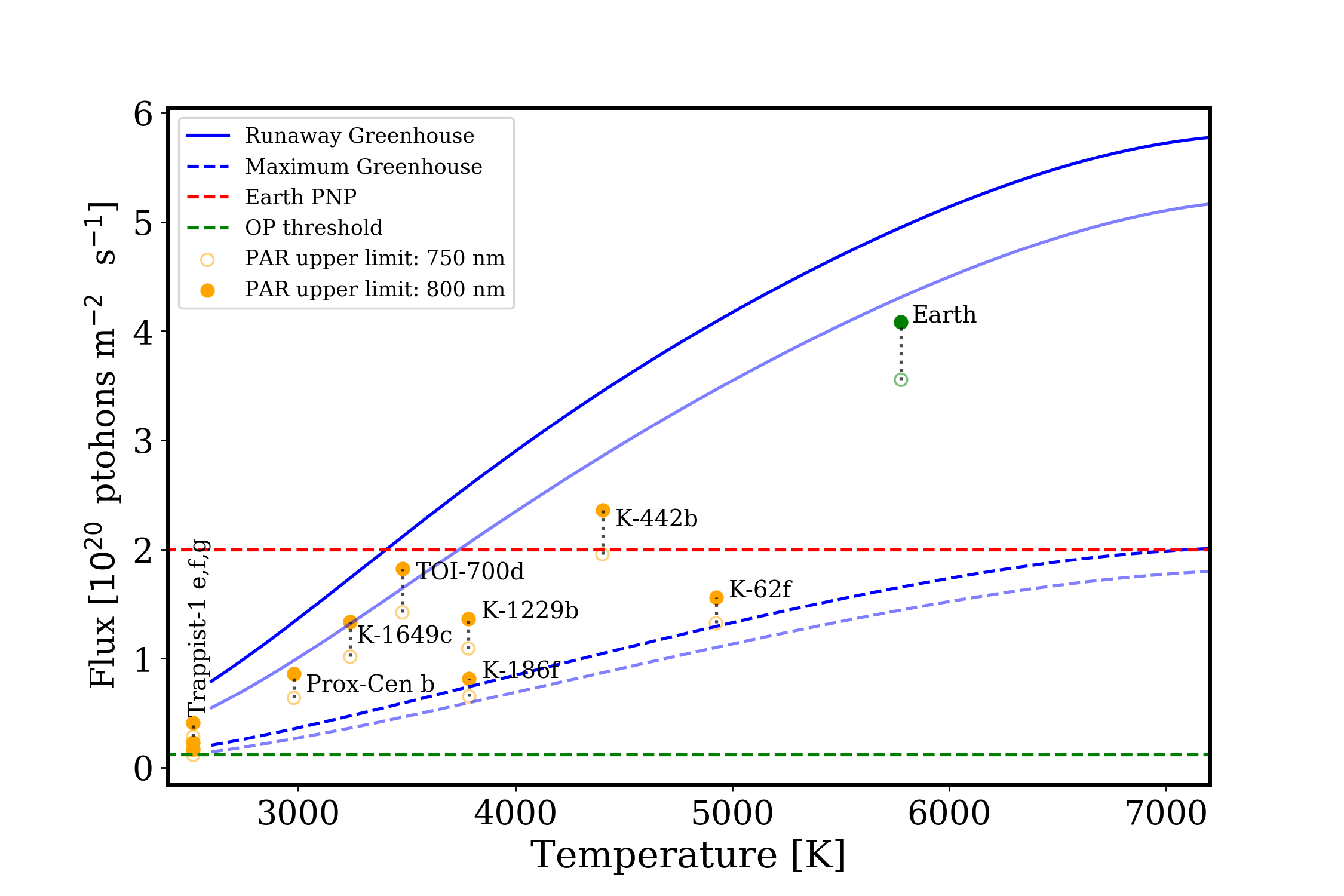}
\caption{Photons flux in two differently defined PAR ranges at the surface of planets at the the two edges of the HZ (dark blue lines for an upper limit of 800 nm and light blue for an upper limit of 750 nm), as a function of the star effective temperature, in units of $10^{20} \, {\rm photons}
\, {\rm s}^{-1} \, {\rm m}^{-2}$ (HZ inner edge: continuous line;
HZ outer edge: dotted line).
The green dot and circle show the photon flux in PAR range on the Earth surface, yellow dots and circles the estimated photon flux on the surface of known Earth analogs (see Table \ref{tab:planets}) respectively with an upper limit for the PAR range of 800 nm (dots) and 750 nm (circles).
The red dotted line shows the average photon flux which is necessary to sustain the Earth biosphere.
The green dotted line shows the typical lower threshold for OP on Earth.}
\label{fig:flux_PAR}
\end{figure*}

As the OP reaction (\ref{eq:OP}) depends on the number of PAR photons rather than the overall energy flux, we focus here of the photon flux
within the PAR range given by the well-known Planck law (at the star surface): 
\begin{equation}
N_{\gamma}^{(0)} (T) \, = \, 
\int_{\lambda_{i n f}}^{\lambda_{\rm sup}} \frac{2 \pi c}{\lambda^4} \,  \frac{1}{e^{\frac{h c}{\lambda k T}}-1} \, {\rm d}  \lambda \,  ,
\label{eq:photon_flux_1}
\end{equation}
where $\lambda_{\rm inf}=400 \, {\rm nm}$ and $\lambda_{\rm sup}=800 \, {\rm nm} .$ 
The Sun emits about 34\% 
of its photon flux in this range.

The PAR photon flux $N_{\gamma}$ at the surface of  
planet at distance $d$ from the parent star is
\begin{equation}
N_{\gamma} (T) \, = \,
\frac{1}{4} \, 
\left( \frac{R}{d} \right)^2 \, 
\int_{\lambda_{i n f}}^{\lambda_{\rm sup}} 
\epsilon (\lambda) \,
\frac{2 \pi c}{\lambda^4} \,  \frac{1}{e^{\frac{h c}{\lambda k T}}-1} \, {\rm d}  \lambda \,  ,
\label{eq:photon_flux_2}
\end{equation}
The factor 1/4 comes from the fact that the stellar radiation intercepted by the planet (with cross section $\pi R_p^2$, where $R_p$ is the planet radius) is averaged on the entire surface.
$R$ is the stellar radius and the quantity $\epsilon (\lambda)$
is the planet atmospheric transmittance, that is the ratio between
the flux at the surface and the incident flux, at given wavelength. 

We computed the  atmospheric transmittance for each planet by using 
the Planetary Spectrum Generator \citep[PSG,][]{Villanueva2015S, Villanueva2018}.
PSG\footnote{Web site: https://psg.gsfc.nasa.gov/} is a publicly available online calculator which allows to
compute (among other quantities) exoplanets transmittances, reflectances and emissivities.
The parameter $\epsilon (\lambda)$ depends on the chemical composition of the atmosphere and on the planet mass (that affects the scale height of the atmosphere), which in turn is 
computed from the planet size (see Table \ref{tab:planets}) and assuming the Earth density (when the mass has not been directly measured). 
While in previous works authors generally adopted a model with no atmosphere  \citep[e.g.][]{Lingam2019a}, here we consider atmospheres with the composition of modern Earth. We checked that results obtained using the likely atmospheric composition of the Archean Earth only differ of a few percents, thus well within the uncertainties of our model. We do not consider clouds in our atmospheric models. 
The trasmittance $\epsilon (\lambda)$ generally increases as a function of the wavelength as a consequence of the dependence of the Rayleigh scattering on this quantity, see also  \cite{Kopparapu2013}. 
For a H$_2$O-dominated atmosphere, the parameter $\epsilon (\lambda)$ shows shallow absorption bands at the longer wavelengths (750-800 nm). 
For a CO$_2$-dominated atmosphere, $\epsilon (\lambda)$ does not show any significant absorption feature in the PAR.

%



Over the PAR range, $\epsilon$ is larger than the corresponding coefficient relative to entire spectrum, as the atmosphere is almost optically thin for the PAR.

We calculate the PAR photon flux for Earth analogs at the HZ edges.
Using equations (\ref{eq:koppa}) and (\ref{eq:photon_flux_2}) 
we obtain the following relation for the photon flux at the planet surface: 
\begin{equation}
N_{\gamma} (T)  = \frac{1}{4} \, 
\left( \frac{R_{\odot}}{\rm AU}  \right)^2 \,
\left( \frac{T_{\odot}}{T} \right)^4 \, S_{\rm eff} (T) \, 
\int_{\lambda_{\rm inf}}^{\lambda_{\rm sup}} 
\frac{2 \pi c}{\lambda^4} \,  \frac{\epsilon (\lambda)}{e^{\frac{h c}{\lambda k T}}-1} \, {\rm d}  \lambda \,  .
\label{eq:qpp}
\end{equation}
Note that this relation does not depend on the stellar radius. 
In Fig.~\ref{fig:flux_PAR} we plot the expected PAR flux for planets at the HZ inner and outer edges, as a function of the host star effective temperature
and compare these quantities with the predictions  
for the sample of confirmed Earth-like planets in the HZ (see Appendix \ref{sec:appA} for details), computed using Eq. (\ref{eq:photon_flux_2}).
The expected PAR fluxes for the two edges of the HZ have been computed by assuming $T_{\rm eq} = 300$ K
and the atmospheric models used in \cite{Kopparapu2013} to define the HZ.
At the HZ inner edge the atmosphere is assumed to be H$_2$O-dominated, at the outer edge the atmosphere is CO$_2$ dominated. 

Both curves have been computed assuming two different PAR upper limits (750 and 800 nm) in order to quantify the PAR photon flux increase due to the PAR extension, see Fig.˜1.
The impact of the extension the PAR to 800 nm is larger for cool stars (the photon flux increases by about 40\% at $T<3000$ K) and less significant for hotter stars (less than 10\% at $T>6000$ K).

The PAR flux is a monotonic function of the star temperature, increasing by about one order of magnitude in the considered temperature range.
This result is at odds with the claim by \cite{Lingam2019a}
who find that the PAR photon flux received on the surface of a habitable Earth-analog as a function of the stellar mass has a maximum at
$M \sim M_{\odot}$, while we see no maximum at $T \sim T_{\odot}$, 
see Fig. \ref{fig:flux_PAR}.
\cite{Lingam2019a} assume indeed a simple model for Earth-analogs,
without taking into account the dependence of the HZ from the climatic models, 
as done  for instance in \cite{Kopparapu2013}.
In Eq. (1) in  \cite{Lingam2019a}, authors implicitly assume $S_{\rm eff}$ to be 1, when relating the orbital distance of the planet to the host star luminosity.
We also computed the expected PAR photon fluxes at the surface of Earth and the exoplanets sample for both definition of the PAR. 

Finally, in Fig. \ref{fig:flux_PAR} we compare our results
with the value of the PAR photon flux incident on the planet that is necessary to sustain the so-called net primary productivity equal to that of the Earth.
According to \cite{field1998primary}, this value is about
$2 \times 10^{20} \, {\rm photons}
\, {\rm m}^{-2} \, {\rm s}^{-1} \, .$
This value is lower than the actual photon flux at the Earth surface
as the biomass production is mostly limited by the 
abundance of nutrients and not by abundance of light. 
We see that all exoplanets in our sample (but Kepler-442b) are well below such a threshold. 
Hence, the PAR photon flux on these exoplanets 
would probably not sustain a biosphere as the one on our planet, if the OP chemical reaction (\ref{eq:OP}) using PAR is the main mechanism to produce biomass.
Kepler-442b stands out as the only other world where 
OP could produce a biosphere theoretically as the one on our planet.
Extending the PAR up to 800 nm would have an important impact, as
the photon flux would increase by about 20 \%. 
Kepler-442b orbits a K-type star with mass 0.61 M$_{\odot}$ and is not tidally locked. 

\input{Table1}

In Fig. \ref{fig:flux_PAR} we also compare 
predictions for Earth-like planets with the 
approximate value of the OP threshold
as observed on Earth.
Indeed, studies of phytoplankton OP have found that 
the minimum irradiance for net positive photosynthesis
is about 2-20 $\mu$mol photon 
$\, {\rm m}^{-2} \, {\rm s}^{-1} \, ,$
see, e.g., \cite{falkowski2013aquatic}.
Phytoplankton OP has been observed also at sea depth where
the photon flux is about 3 orders of magnitude lower 
\citep{raven2000, Wolstencroft2002},
but in this regime the contribution to the biosphere is
negligible.
Generally speaking, low mass stars provide enough photons to start the OP on planets in their HZ,
however 
the outer edge of the HZ around cool stars with 
$T \lesssim 3000 \, $ K lies around this minimum threshold.
This value is not a sharp 
and universal threshold for OP, but it is indicative of a regime where likely the low rate of the primary production could not support major food chains, \citep{raven2000}. 
Our results are in agreement with recent findings about the low net primary productivity 
of M-dwarfs exoplanets \citep{Ritchie2018I}.
In particular, two planets from the Trappist-1 system
\citep{Gillon2017} fall short of this threshold.
This is in agreement with results from \cite{Lehmer2018} who found that biospheres on exoplanets around Trappist-1 are potentially light-limited. 


\section{Exergy and exergetic efficiency of stellar radiation}

In this section we estimate the efficiency 
of the PAR radiation driving OP as a function of the host-star temperature by means of the notion of exergy. 
Exergy can be defined as the maximum useful work obtainable from the considered system in given environmental conditions, see, e.g., 
\cite{Petela2008, Ptasinski2016efficiency}.
In other words, exergy is a measure of the quality of energy \citep{AUSTBO2014391}.
Living organisms are dissipative structures away from thermodynamic equilibrium with the environment
thanks to the constant input of exergy 
stellar radiation.
%

Several authors have presented 
energetic and exergetic analyses of the OP 
on Earth, see, e.g, \cite{Petela2008, Silva2015,  Delgado-Bonal_2017}.
However, it is not immediate to extend their results to the possible OP on Earth-like exoplanets, as both the environmental conditions and the specific properties of the OP could be very different.

Recently, \cite{Scharf2019} applied the notion of exergy to the input stellar radiation to investigate 
the OP efficiency as a function of the star effective temperature, considering the entire spectrum. 
Generally speaking, the exergy flux $E{\rm x}$ of the radiation with flux $F$ investing a system (planet) with surface temperature $T_0$
can be written as \citep{Candau2003}:
\begin{equation}
E{\rm x} \, = \, F - F_0 \, - T_0 \, (S
- S_0 ) \, ,
\label{eq:exergy1}
\end{equation}
where $F_0$ is the flux emitted by the environment, 
$S$ and $S_0$ are entropy fluxes relative to $F$ and $F_0$ energy fluxes.
Flux and entropy are evaluated in a generic wavelength range.
When we consider the radiation full wavelength range, we obtain the following expression  \citep{Delgado-Bonal_2017}: 
\begin{equation}
E{\rm x} \, =
\, \sigma \, \left(T_{\rm }^4- \frac{4}{3} T_0 \, T_{\rm }^3 + \frac{1}{3} T_0^4\right) \, .
\label{eq:exergy2}
\end{equation}
This is the the maximum obtainable work from the whole radiation.
\cite{Scharf2019} used this expression to show that the efficiency of the radiation investing a planet
increases with the stellar temperature,
at given planet surface temperature.
In the following, as plants do not use the whole incident radiation, we consider the exergy in the PAR range.
We also note that Eq. (\ref{eq:exergy2}) does not take into account the dependency from the star-planet separation.

By considering the well-known relation between energy flux and entropy flux for a blackbody radiation, we can recast Eq. (\ref{eq:exergy1}) as follows:
\begin{equation}
E{\rm x} \,  = \left(1 - \frac{4}{3} \frac{T_0}{T}\right) \, F + \frac{F_0}{3} \, , 
\label{eq:exergy3}
\end{equation}
see \cite{Petela_1964}.
We use this expression to compute the PAR exergy flux at the planet surface.
Let $F^{(0)}_{\rm PAR} (T)$ be the star PAR flux at its surface:
\begin{equation}
F^{(0)}_{\rm PAR} (T)  = 
\int_{\lambda_{\rm inf}}^{\lambda_{\rm sup}} 
 \frac{2 \pi h c^2}{\lambda^5} \,  \frac{1}{e^{\frac{h c}{\lambda k T}}-1} \, {\rm d}  \lambda \,  , 
\end{equation} 
and $F_{\rm PAR}$ the stellar flux at the planet surface in HZ,  using Eq. (\ref{eq:koppa}), is
\begin{equation}
F_{\rm PAR} (T)  = \frac{1}{4} \,\left( \frac{R_{\odot}}{\rm AU} 
\right)^2 \left( \frac{T_{\odot}}{T}
\right)^4 \,S_{\rm eff} \int_{\lambda_{\rm inf}}^{\lambda_{\rm sup}} \frac{2 \pi h c^2}{\lambda^5} \,  \frac{\epsilon \, (\lambda)}{e^{\frac{h c}{\lambda k T}}-1} \, {\rm d}  \lambda \,  , 
\label{eq:energy_flux_1}
\end{equation} 
and by replacing this relation in Eq. (\ref{eq:exergy3}), we obtain
\begin{equation}
E{\rm x}_{\rm PAR} = \left( 1 - \frac{4}{3} \frac{T_0}{T} \right) 
 \, F_{ \rm PAR}  + \frac{F_{0, \rm PAR}}{3} \, ,
\end{equation}
%
%
where the energy flux from the planet $F_{0, \rm PAR}$ is computed adopting a blackbody radiation, with $T_0 = 293$ K.
We note that the contribution of the term $F_{0}$ is negligible in the PAR range, as $T_0 \ll T$.

\begin{figure}
\centering
\includegraphics[width=0.5\textwidth]{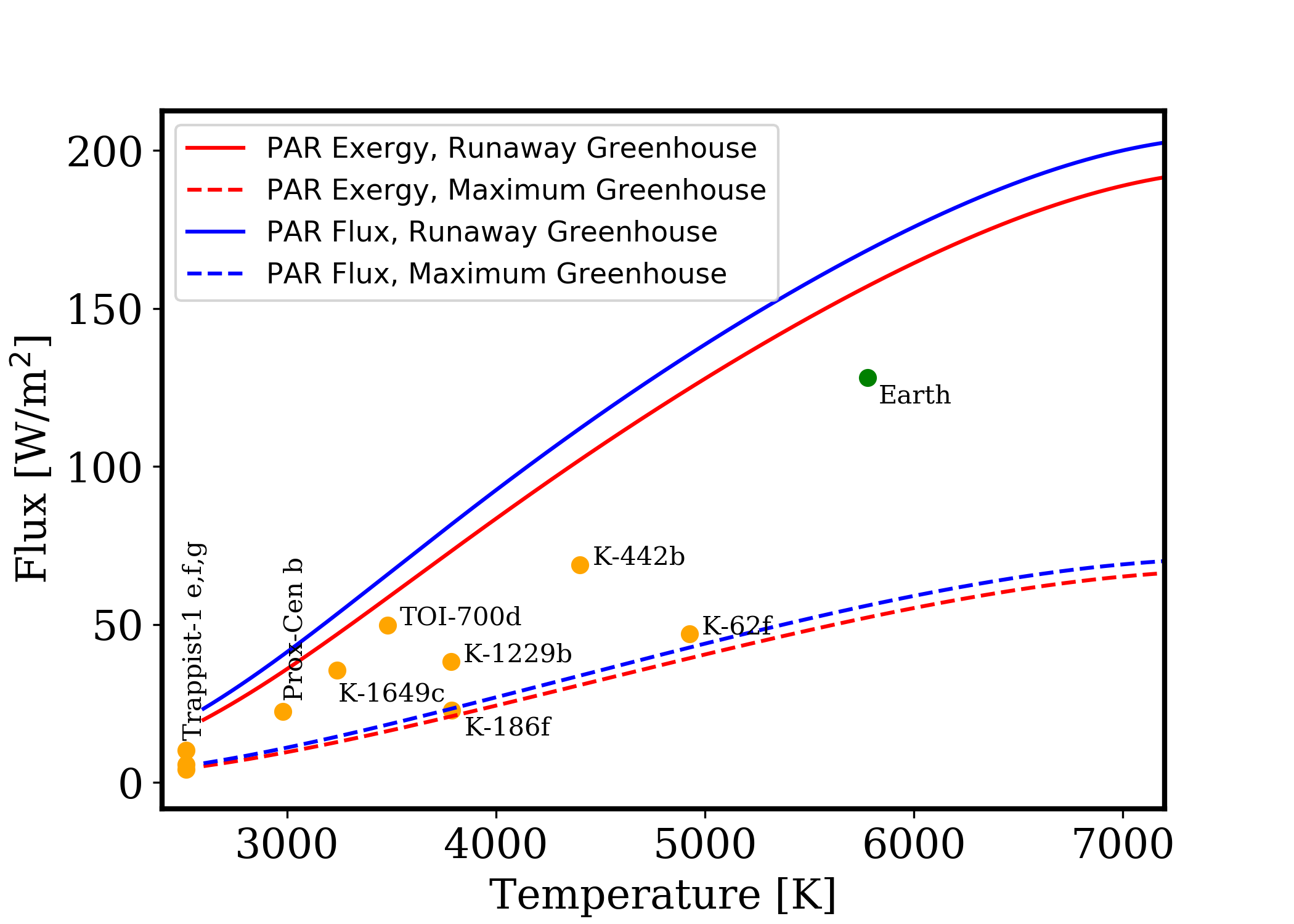}
\caption{Exergy flux (red lines) and the total energy flux (blue lines) within the PAR range (400-800 nm)
as a function of the host star effective temperature.
Continuous lines represent the values at the surface of planets 
at the HZ inner edges, dotted lines for the HZ and outer edges. 
Orange dots represent the estimated exergy flux on the surface for planets from the sample in Table \ref{tab:planets}.}
\label{fig:exegy_par_hz_vs_flux}
\end{figure}

In Fig. \ref{fig:exegy_par_hz_vs_flux} we plot the energy and exergy fluxes for the inner and outer HZ edges, together with the average flux and exergy values for the planets in our sample.
As expected, both energy and exergy flux increase with the effective star temperature and with decreasing distance to the stars.

The exergy flux is always lower than the energy flux: the main reason is that the entropy content of the radiation reduces the ability to extract useful work the radiation field.
We also note that the exergy flux on a planet depends also on its distance from the star. This is in contrast with the findings by \cite{Scharf2019}, which did not consider any dependence on the distance from the star.

In Fig. \ref{fig:exegy_par_hz_vs_flux}
we also compare the exergy flux values for Earth and the planets sample 
in Table \ref{tab:planets}. 
There is no known HZ terrestrial planet with exergy values close to the Earth value.

We now discuss the exergetic efficiency of the star radiation within the PAR. A common definition of the exergetic efficiency 
is the ratio between the
 exergy within the PAR range
and the incoming radiative energy in all wavelengths , see, e.g., \cite{Delgado-Bonal_2017}. Therefore, 
the exergetic efficiency can be defined as 
\begin{equation}
\eta_{E{\rm x}, {\rm PAR}} \,
= \, \frac{E{\rm x}}{F} \, .
\label{eq:exergy_efficiency}    
\end{equation}
This definition differs from the one used by \cite{Scharf2019} as it considers the thermodynamic quantities evaluated at the planet
location (not at the star surface)
and considers the exergy over the PAR wavelength range (not the entire spectrum). 

As noted above, we can neglect the contribution 
from the planet flux $F_{0, \rm PAR}$. Hence, the expression for the exergetic efficiency reads:  
\begin{equation}
\eta_{E{\rm x}, {\rm PAR}} \, \simeq \left(1-\frac{4}{3} \frac{T_{0}}{T}\right) \,\frac{F_{PAR}(T)}{F_{tot}(T)} \,
\label{eq:eff2}
\end{equation}
where
\begin{equation}
    F_{\rm tot}(T) = \frac{1}{4} \,\left( \frac{R_{\odot}}{\rm AU} 
\right)^2 \left( \frac{T_{\odot}}{T}
\right)^4 \,S_{\rm eff} \int_{{\lambda_{min }}}^{\lambda_{max}} 
 \frac{2 \pi h c^2}{\lambda^5} \,  \frac{\epsilon(\lambda)}{e^{\frac{h c}{\lambda k T}}-1} \, {\rm d}  \lambda \, , 
\end{equation}
where $\lambda_{\rm min} = 0.25 \mu m $ and $ \lambda_{\rm max} = 10 \mu$m. We set $\lambda_{\rm min} = 0.25 \mu$m  as the computed transmittance is very close to zero at shorter wavelengths for the considered atmospheric models.
By integrating the stellar blackbody spectrum up to $10 \mu m$, we are committing an error in the computed total stellar flux $F_{\rm tot}(T)$ of less than a percent in the considered temperature range (2600 - 7200 K).

\begin{figure}
\centering
\includegraphics[width=0.5\textwidth]{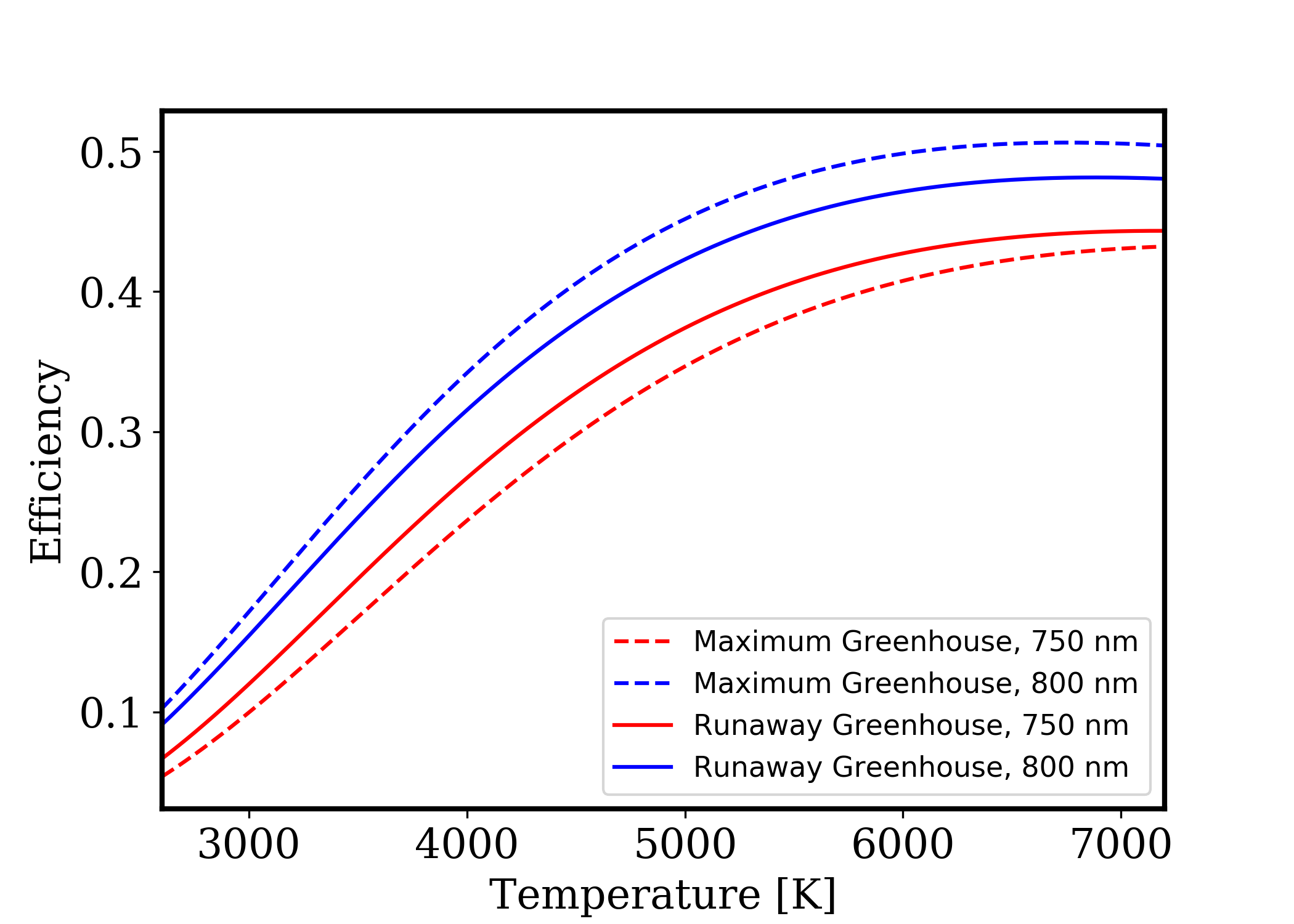}
\caption{Exergetic efficiency versus the host star effective temperature for planets at the edges of the HZ for the two different PAR ranges.}
\label{fig:efficiency}
\end{figure}

If we consider worlds with surface temperature close to that of our planet, in first approximation $\eta_{E{\rm x}, {\rm PAR}} $ only depends on the host star effective temperature. 
In particular, it does not depend on the distance of the planet from the host star.
The PAR exergetic efficiency is shown in Fig. \ref{fig:efficiency}, for the atmospheric models at the HZ edges and the two PAR ranges considered in this work. 
The efficiency is a monotonic increasing function of the host star temperature (as already noted by \cite{Scharf2019} considering the whole radiation spectrum), but it is nearly constant for stars hotter than the Sun. 
When the PAR is extended, the exergetic efficiency increases for both atmospheric models.
We also note that the maximum greenhouse atmospheric model allows a larger exergetic efficiency than runaway greenhouse atmospheric model as a consequence of the H$_2$O absorption feature.


\section{Conclusions}

The rapid progress in exoplanet science
has led us closer to the detection of biosignatures and possibly technosignatures on other worlds.
Planets are being discovered 
at increasing pace around nearby stars
(see \cite{Kaltenegger2017ARA&A..55..433K} for a recent review),
towards the Galactic bulge \citep{Tsapras2018Geosc} and up to the nearby galaxies \citep{Covone2000, An2004ApJ}.
In the next few years 
the small sample of rocky planets known today will 
be much enlarged and we will start to  assess  their habitability via spectroscopic observations \citep{Seager2010ARA&A..48..631S}.
The concept of habitability is complex 
and multifaceted. Recently, \cite{Scharf2019} 
has pointed that exergy of radiation should also be taken into account when evaluating the possibility to find a robust biosphere on other worlds. 
In this paper we further addressed this topic 
by estimating the PAR photon flux and its exergetic efficiency and compared 
estimates for the known Earth-analogs in their HZ (see Table \ref{tab:planets}).


Earth is by large the rocky planet with the largest PAR photon flux
and with the highest exergetic efficiency.
However, we also find that Kepler-442b receives a PAR photon flux  slightly larger than the one necessary to sustain a large biosphere,
similar to the Earth biosphere. 
So, it is likely that a Kepler-442b
biosphere would not be light-limited. 
It is also worthy to notice that Kepler-442 is not tidally locked and orbits a K-type star. This makes this planet a promising target for
search of biosignatures, as \cite{Cuntz2016ApJ} have shown that K-type stars provide favourable circumstellar environment for life.
Moreover K-dwarf stars appear to present a biosignature detectability advantage \citep{Arney2019ApJ}.

However, we should bear in mind that biomass production on Earth is not limited by the quantity neither the quality of the incoming radiation, but rather by the availability of nutrients.
For instance, \cite{Lin2016} found that 
in ocean phytoplankton populations 
about 60\% of the absorbed PAR solar energy is dissipated as heat. 
Generally, phytoplankton operate at a much lower photosynthetic efficiency than they are potentially capable of achieving, just because in most situations light is a very abundant resource on Earth.
Moreover, OP does not respond linearly to the input photon flux, see \cite{Ritchie2018I}.
For these reasons, it is not immediate to draw consequences on the amount of biomass produced from the estimated PAR photon flux and its exergy content. 
Exoplanets with lower values of these quantities could host a biosphere comparable with the one on our planet.

In the present work, we focused on the effect of incoming stellar radiation and did not consider the complex details of OP process.
Our estimates have to be considered upper limits as, for instance, we did not take into account the exergy destruction that occurs as consequence of biological conversion taking place after the light harvesting, in the leaf
transpiration and  metabolism.
We also neglected other effects that reduce the efficiency 
of the radiation: for instance, the atmosphere absorption also changes
the radiation spectrum, increasing its entropy content, hence reducing the useful work.
Next models should also consider cloud coverage which reduces the radiation available for the OP. 

We focused on oxygenic photosynthesisdriven by optical light, up to 800 nm. However, other possible solutions could be found in Nature to convert radiation into organic compounds. For instance, \cite{Takizawa2017Nat} discuss the possible evolution 
of OP organisms harvesting near-infrared light on planets around red stars.
Indeed, it is possible to consider OP reactions 
with three  or four photons, instead of a two-photon reaction, with excitation wavelengths  1050 and 1400 nm, respectively \citep{Takizawa2017Nat}. 
Other authors studied OP driven by NIR light, see 
\cite{Wolstencroft2002}, \cite{Kiang_2007AsBio...7..252K} and \cite{Lehmer2018}.
In future works, it will be important to extend the present analysis to 
OP using NIR light as well as to the case of anoxygenic photosynthesis,
where hydrogen donors different from water are involved. For instance, on Earth 'purple bacteria' use H$_2$S instead of water. 
Anoxygenic photosynthesis organisms are capable to harvest light at longer wavelength.
This could be a considerable evolutionary advantage for these organisms
on planets around cool stars, see \cite{Bains2014}.
Indeed, \cite{Ritchie2018I} has estimated that larger amounts of light 
(about $10^3 \, \mu$mol  photons $\, {\rm s}^{-1} \, {\rm m}^{-2}$)
would be available for anoxygenic photosynthesis (using bacteriochlorophyll)
in the wavelength range between 350 and 1100 nm.

\section*{Acknowledgements}

We thank Simona Carfagna and Ravi K. Kopparapu for many insightful discussions.
We thank the anonymous referee for helpful comments on the manuscript. 
This research has made use of the Planetary Spectrum Generator (NASA) and the Planet Habitability Laboratory website hosted by the University of Puerto Rico at Arecibo.

\section*{Data availability}

No new data were generated or analysed in support of this research. All data used in this paper were obtained from the cited literature.

\bibliographystyle{mnras}
\bibliography{biblio_exobiology} 




\appendix

\section{Confirmed HZ Earth-like planets}
\label{sec:appA}

We collected a conservative sample of confirmed Earth-like exoplanets, with measured mass and orbits in the HZ (as defined in  \cite{Kopparapu2013}). 
The sample includes 10 exoplanets. 
When the radius is known (for exoplanets discovered  via the photometric transit technique),
we selected systems with $R<1.5R_\oplus$ and confidently assume their rocky composition. 
We listed the characteristics of these exoplanets and the properties of their host stars in Table \ref{tab:planets}.

We used the Planet Habitability Laboratory website\footnote{http://phl.upr.edu/projects/habitable-exoplanets-catalog},  which is constantly upgraded with the latest discovered potential habitable exoplanets.

The dataset comprehends three out the seven planets orbiting Trappist-1 \citep{Gillon2017}. 
The habitability of the Trappist-1 planets has long been discussed as it is likely that 
some of the planets might have undergone runaway greenhouse effect (such as Venus) or experienced the impetuous activity of their $\approx 0.08R_\odot$ star. 
Kepler-186f, one of the first Earth-like planets ever discovered orbiting the HZ of a Sun-like star, is the outermost of a system of five Earth-sized planets \citep{Quintana2014}. 

There are other four HZ exoplanets revolving stars in the Kepler catalog: Kepler-62f \citep{Borucki2013}, Kepler-442b \citep{Torres2015}, Kepler-1229b \citep{Torres2017} and Kepler 1649c \citep{Angelo2017} that spread a wide range in host star temperatures. 

TOI700-d is a recent addition to the sample
and the only one discovered with data from the Transiting Exoplanet Survey Satellite [\citep{Gilbert2020}, \citep{Rodriguez2020}]. 
Its radius is  $1.1 R_\oplus$
and orbits a quiet M-dwarf star at 31 pc from the Sun.  
Lastly, we included  the exoplanet Proxima Centauri b. 
It has been discovered with the radial velocity method \citep{Anglada-Escude2016}, yielding only a lower limit for its mass of $1.27 M_\oplus$
(see also \citep{Jenkins2019}). 
It is worth to notice that Proxima Centauri shows a significant activity \citep{Vida2019}, making the
effective habitability of the host  exoplanet highly uncertain. 

We remark that this sample is not complete. A complete census of Earth-like planets in the HZ would require the use of a combination of search strategies, including next imaging facilities which could more efficiently discover such planets in the HZ of hotter stars \citep{Arney2018arXiv}.


\bsp	
\label{lastpage}
\end{document}

%% file: Table1.tex
\begin{table*}
\centering
\begin{tabular}{|l|l|l|l|l|l|l|l|l|}
\hline  Star &  $T_{S}$ (K)& $\mathrm{R}_{star} \, ({R}_{\odot})$ & Planet & $R_{\rm p} \, (R_{\oplus})$ & $a \, (\mathrm{AU})$ & $\mathrm{T}_{\text {eq }}(\mathrm{K})$  & reference \\
\hline Trappist-1  & $2559 \pm 50$ & $0.121 \pm 0.003$ & Trappist-1e & $0.772 \pm 0.077$ & $0.02817_{-0.00087}^{+0.00083}$ & $251.3 \pm 4.9$ &  \cite{Gillon2017}\\
\hline Trappist-1 & $2559 \pm 50$ & $0.121 \pm 0.003$ & Trappist-1f & $0.934 \pm 0.079$  & $0.0371\pm0.0011$ & $219.0 \pm 4.2$ & \cite{Gillon2017} \\
\hline Trappist-1 & $2559 \pm 50$ & $0.121 \pm 0.003$ & Trappist-1g & $1.148 \pm 0.097$  & $0.0451\pm0.0014$ & $198.6 \pm 3.8$ & \cite{Gillon2017} \\

\hline Kepler-186  &  $3788 \pm 54$ &  $0.472_{-0.05}^{+0.17}$ & $\mathrm{Kepler-186 f}$ & $1.11_{-0.13}^{+0.14}$  & $0.356\pm 0.048$ & $202\pm30 \, (*)$  & \cite{Quintana2014} \\

\hline Kepler-62  &$4925 \pm 70$  & $0.64 \pm 0.02$ & $\mathrm{Kepler-62f}$ & $1.41\pm 0.07$ & $0.718 \pm 0.007$ & $208 \pm 11$ & \cite{Borucki2013}\\

\hline  Kepler-442 &  $4402 \pm 100$ & $0.598 \pm 0.023$ & $\mathrm{Kepler-442b}$ & $1.34_{-0.18}^{+0.11}$ &  $0.409_{-0.20}^{+0.06}$ & $146_{-22}^{+80}$ & \cite{Torres2015} \\

\hline 
Kepler-1229  &  $3577 \pm 58$ &  $0.510 \pm 0.029$ & $\mathrm{Kepler-1229b}$ & $1.34_{-0.14}^{+0.01}$ &  $0.3006_{-0.009}^{+0.007}$ & $212_{-17}^{+19}$ & \cite{Torres2017} \\

\hline Kepler-1649  &  $ 3240\pm61 $ &   $0.2317\pm0.0049 $ & $\mathrm{Kepler-1649c}$ & $ 1.06\pm0.12 $  & $ 0.0827\pm0.0017$ & $ 234\pm20 $ & \cite{Angelo2017} \\


\hline TOI-700  &  $3480 \pm 135$ &  $0.420 \pm 0.031$ & $\mathrm{TOI - 700d}$ & $1.144_{-0.061}^{+0.062}$ & $0.1633_{-0.0026}^{+0.0027}$ & $268\pm7$ & \cite{Rodriguez2020} \\

\hline Proxima Cen  & $ 2980 \pm 80$ &  $0.146 \pm 0.007$ & Proxima Cen b & $1.27\pm0.19 $ & $0.0485\pm0.05$ & $230\pm10$ & \cite{Anglada-Escude2016}\\
\hline

\end{tabular}
\caption{Parameters of the known Earth analog planets in the HZ and their host stars. Equilibrium temperature values with $*$ have been derived in this work. For Proxima Centauri b the esteem of the mass is given since the planet is probably not a transiting one. }
\label{tab:planets}
\end{table*}


%% file: main.bbl
\begin{thebibliography}{}
\makeatletter
\relax
\def\mn@urlcharsother{\let\do\@makeother \do\$\do\&\do\#\do\^\do\_\do\%\do\~}
\def\mn@doi{\begingroup\mn@urlcharsother \@ifnextchar [ {\mn@doi@}
  {\mn@doi@[]}}
\def\mn@doi@[#1]#2{\def\@tempa{#1}\ifx\@tempa\@empty \href
  {http://dx.doi.org/#2} {doi:#2}\else \href {http://dx.doi.org/#2} {#1}\fi
  \endgroup}
\def\mn@eprint#1#2{\mn@eprint@#1:#2::\@nil}
\def\mn@eprint@arXiv#1{\href {http://arxiv.org/abs/#1} {{\tt arXiv:#1}}}
\def\mn@eprint@dblp#1{\href {http://dblp.uni-trier.de/rec/bibtex/#1.xml}
  {dblp:#1}}
\def\mn@eprint@#1:#2:#3:#4\@nil{\def\@tempa {#1}\def\@tempb {#2}\def\@tempc
  {#3}\ifx \@tempc \@empty \let \@tempc \@tempb \let \@tempb \@tempa \fi \ifx
  \@tempb \@empty \def\@tempb {arXiv}\fi \@ifundefined
  {mn@eprint@\@tempb}{\@tempb:\@tempc}{\expandafter \expandafter \csname
  mn@eprint@\@tempb\endcsname \expandafter{\@tempc}}}

\bibitem[\protect\citeauthoryear{{An} et~al.,}{{An} et~al.}{2004}]{An2004ApJ}
{An} J.~H.,  et~al., 2004, \mn@doi [\apj] {10.1086/380820}, \href
  {https://ui.adsabs.harvard.edu/abs/2004ApJ...601..845A} {601, 845}

\bibitem[\protect\citeauthoryear{Anbar et~al.,}{Anbar et~al.}{2007}]{Anbar2007}
Anbar A.~D.,  et~al., 2007, \mn@doi [Science] {10.1126/science.1140325}, 317,
  1903

\bibitem[\protect\citeauthoryear{{Angelo} et~al.,}{{Angelo}
  et~al.}{2017}]{Angelo2017}
{Angelo} I.,  et~al., 2017, \mn@doi [\aj] {10.3847/1538-3881/aa615f}, \href
  {https://ui.adsabs.harvard.edu/abs/2017AJ....153..162A} {153, 162}

\bibitem[\protect\citeauthoryear{{Anglada-Escud{\'e}}
  et~al.,}{{Anglada-Escud{\'e}} et~al.}{2016}]{Anglada-Escude2016}
{Anglada-Escud{\'e}} G.,  et~al., 2016, \mn@doi [\nat] {10.1038/nature19106},
  \href {https://ui.adsabs.harvard.edu/abs/2016Natur.536..437A} {536, 437}

\bibitem[\protect\citeauthoryear{{Arney}}{{Arney}}{2019}]{Arney2019ApJ}
{Arney} G.~N.,  2019, \mn@doi [\apjl] {10.3847/2041-8213/ab0651}, \href
  {https://ui.adsabs.harvard.edu/abs/2019ApJ...873L...7A} {873, L7}

\bibitem[\protect\citeauthoryear{{Arney} et~al.,}{{Arney}
  et~al.}{2018}]{Arney2018arXiv}
{Arney} G.,  et~al., 2018, arXiv e-prints, \href
  {https://ui.adsabs.harvard.edu/abs/2018arXiv180302926A} {p. arXiv:1803.02926}

\bibitem[\protect\citeauthoryear{{Austbø}, {Løvseth}  \&
  {Gundersen}}{{Austbø} et~al.}{2014}]{AUSTBO2014391}
{Austbø} B.,  {Løvseth} S.~W.,   {Gundersen} T.,  2014, \mn@doi [Computers \&
  Chemical Engineering] {https://doi.org/10.1016/j.compchemeng.2014.09.010},
  71, 391

\bibitem[\protect\citeauthoryear{{Bains}, {Seager}  \& {Zsom}}{{Bains}
  et~al.}{2014}]{Bains2014}
{Bains} W.,  {Seager} S.,   {Zsom} A.,  2014, \mn@doi [Life]
  {10.3390/life4040716}, 4, 716

\bibitem[\protect\citeauthoryear{Benner, Ricardo  \& Carrigan}{Benner
  et~al.}{2004}]{Benner_2004}
Benner S.~A.,  Ricardo A.,   Carrigan M.~A.,  2004, \mn@doi [Current Opinion in
  Chemical Biology] {https://doi.org/10.1016/j.cbpa.2004.10.003}, 8, 672

\bibitem[\protect\citeauthoryear{{Borucki} et~al.,}{{Borucki}
  et~al.}{2013}]{Borucki2013}
{Borucki} W.~J.,  et~al., 2013, \mn@doi [Science] {10.1126/science.1234702},
  \href {https://ui.adsabs.harvard.edu/abs/2013Sci...340..587B} {340, 587}

\bibitem[\protect\citeauthoryear{{Candau}}{{Candau}}{2003}]{Candau2003}
{Candau} Y.,  2003, \mn@doi [Solar Energy] {10.1016/j.solener.2003.07.012},
  \href {https://ui.adsabs.harvard.edu/abs/2003SoEn...75..241C} {75, 241}

\bibitem[\protect\citeauthoryear{Cardona}{Cardona}{2018}]{Cardona2018}
Cardona T.,  2018, Heliyon, 4

\bibitem[\protect\citeauthoryear{Carvalho, Silva  \& Baptista}{Carvalho
  et~al.}{2011}]{Carvalho2011_UV}
Carvalho A.,  Silva S.,   Baptista J. e.~a.,  2011, \mn@doi [Appl Microbiol
  Biotechnol] {10.1007/s00253-010-3047-8}, 89, 1275–1288

\bibitem[\protect\citeauthoryear{{Catling}, {Glein}, {Zahnle}  \&
  {McKay}}{{Catling} et~al.}{2005}]{Catling2005}
{Catling} D.~C.,  {Glein} C.~R.,  {Zahnle} K.~J.,   {McKay} C.~P.,  2005,
  \mn@doi [Astrobiology] {10.1089/ast.2005.5.415}, \href
  {https://ui.adsabs.harvard.edu/abs/2005AsBio...5..415C} {5, 415}

\bibitem[\protect\citeauthoryear{Chen, Schliep, Willows, Cai, Neilan  \&
  Scheer}{Chen et~al.}{2010}]{Chen2010Science}
Chen M.,  Schliep M.,  Willows R.~D.,  Cai Z.-L.,  Neilan B.~A.,   Scheer H.,
  2010, \mn@doi [Science] {10.1126/science.1191127}, 329, 1318

\bibitem[\protect\citeauthoryear{Claudi et~al.,}{Claudi
  et~al.}{2021}]{Claudi-life11010010}
Claudi R.,  et~al., 2021, \mn@doi [Life] {10.3390/life11010010}, 11

\bibitem[\protect\citeauthoryear{{Covone}, {de Ritis}, {Dominik}  \&
  {Marino}}{{Covone} et~al.}{2000}]{Covone2000}
{Covone} G.,  {de Ritis} R.,  {Dominik} M.,   {Marino} A.~A.,  2000, \aap,
  \href {https://ui.adsabs.harvard.edu/abs/2000A&A...357..816C} {357, 816}

\bibitem[\protect\citeauthoryear{{Cuntz} \& {Guinan}}{{Cuntz} \&
  {Guinan}}{2016}]{Cuntz2016ApJ}
{Cuntz} M.,  {Guinan} E.~F.,  2016, \mn@doi [\apj]
  {10.3847/0004-637X/827/1/79}, \href
  {https://ui.adsabs.harvard.edu/abs/2016ApJ...827...79C} {827, 79}

\bibitem[\protect\citeauthoryear{{Delgado-Bonal}}{{Delgado-Bonal}}{2017}]{Delgado-Bonal_2017}
{Delgado-Bonal} A.,  2017, \mn@doi [Scientific Reports]
  {10.1038/s41598-017-01622-6}, \href
  {https://ui.adsabs.harvard.edu/abs/2017NatSR...7.1642D} {7, 1642}

\bibitem[\protect\citeauthoryear{Dole}{Dole}{1964}]{dole1964habitable}
Dole S.,  1964, Habitable Planets for Man.
Blaisdell book on the pure and applied sciences, Blaisdell Publishing Company,
  \url {https://books.google.it/books?id=-z9AAAAAIAAJ}

\bibitem[\protect\citeauthoryear{Falkowski \& Raven}{Falkowski \&
  Raven}{2013}]{falkowski2013aquatic}
Falkowski P.,  Raven J.,  2013, Aquatic Photosynthesis: Second Edition.
Princeton University Press, \url
  {https://books.google.it/books?id=kUCrAQAAQBAJ}

\bibitem[\protect\citeauthoryear{Field, Behrenfeld, Randerson  \&
  Falkowski}{Field et~al.}{1998}]{field1998primary}
Field C.~B.,  Behrenfeld M.~J.,  Randerson J.~T.,   Falkowski P.,  1998,
  science, 281, 237

\bibitem[\protect\citeauthoryear{{Fischer}, {Hemp}  \& {Johnson}}{{Fischer}
  et~al.}{2016}]{Fischer2016AREPS}
{Fischer} W.~W.,  {Hemp} J.,   {Johnson} J.~E.,  2016, \mn@doi [Annual Review
  of Earth and Planetary Sciences] {10.1146/annurev-earth-060313-054810}, \href
  {https://ui.adsabs.harvard.edu/abs/2016AREPS..44..647F} {44, 647}

\bibitem[\protect\citeauthoryear{{Gilbert} et~al.,}{{Gilbert}
  et~al.}{2020}]{Gilbert2020}
{Gilbert} E.~A.,  et~al., 2020, \mn@doi [\aj] {10.3847/1538-3881/aba4b2}, \href
  {https://ui.adsabs.harvard.edu/abs/2020AJ....160..116G} {160, 116}

\bibitem[\protect\citeauthoryear{{Gillon} et~al.,}{{Gillon}
  et~al.}{2017}]{Gillon2017}
{Gillon} M.,  et~al., 2017, \mn@doi [\nat] {10.1038/nature21360}, \href
  {https://ui.adsabs.harvard.edu/abs/2017Natur.542..456G} {542, 456}

\bibitem[\protect\citeauthoryear{{Jenkins} et~al.,}{{Jenkins}
  et~al.}{2019}]{Jenkins2019}
{Jenkins} J.~S.,  et~al., 2019, \mn@doi [\mnras] {10.1093/mnras/stz1268}, \href
  {https://ui.adsabs.harvard.edu/abs/2019MNRAS.487..268J} {487, 268}

\bibitem[\protect\citeauthoryear{Jorgensen \& Svirezhev}{Jorgensen \&
  Svirezhev}{2004}]{jorgensen2004towards}
Jorgensen S.,  Svirezhev Y.,  2004, Towards a Thermodynamic Theory for
  Ecological Systems.
Elsevier Science, \url {https://books.google.it/books?id=5j7TMmhQbPoC}

\bibitem[\protect\citeauthoryear{{Kaltenegger}}{{Kaltenegger}}{2017}]{Kaltenegger2017ARA&A..55..433K}
{Kaltenegger} L.,  2017, \mn@doi [\araa] {10.1146/annurev-astro-082214-122238},
  \href {https://ui.adsabs.harvard.edu/abs/2017ARA&A..55..433K} {55, 433}

\bibitem[\protect\citeauthoryear{Kasting, Whitmire  \& Reynolds}{Kasting
  et~al.}{1993}]{kasting1993habitable}
Kasting J.~F.,  Whitmire D.~P.,   Reynolds R.~T.,  1993, Icarus, 101, 108

\bibitem[\protect\citeauthoryear{{Kiang}, {Siefert}, {Govindjee}  \&
  {Blankenship}}{{Kiang} et~al.}{2007a}]{Kiang_2007_AsBio...7..222K}
{Kiang} N.~Y.,  {Siefert} J.,  {Govindjee}  {Blankenship} R.~E.,  2007a,
  \mn@doi [Astrobiology] {10.1089/ast.2006.0105}, \href
  {https://ui.adsabs.harvard.edu/abs/2007AsBio...7..222K} {7, 222}

\bibitem[\protect\citeauthoryear{{Kiang} et~al.,}{{Kiang}
  et~al.}{2007b}]{Kiang_2007AsBio...7..252K}
{Kiang} N.~Y.,  et~al., 2007b, \mn@doi [Astrobiology] {10.1089/ast.2006.0108},
  \href {https://ui.adsabs.harvard.edu/abs/2007AsBio...7..252K} {7, 252}

\bibitem[\protect\citeauthoryear{{Kiang}, {Domagal-Goldman}, {Parenteau},
  {Catling}, {Fujii}, {Meadows}, {Schwieterman}  \& {Walker}}{{Kiang}
  et~al.}{2018}]{Kiang_2018AsBio}
{Kiang} N.~Y.,  {Domagal-Goldman} S.,  {Parenteau} M.~N.,  {Catling} D.~C.,
  {Fujii} Y.,  {Meadows} V.~S.,  {Schwieterman} E.~W.,   {Walker} S.~I.,  2018,
  \mn@doi [Astrobiology] {10.1089/ast.2018.1862}, \href
  {https://ui.adsabs.harvard.edu/abs/2018AsBio..18..619K} {18, 619}

\bibitem[\protect\citeauthoryear{{Kopparapu} et~al.,}{{Kopparapu}
  et~al.}{2013}]{Kopparapu2013}
{Kopparapu} R.~K.,  et~al., 2013, \mn@doi [\apj] {10.1088/0004-637X/765/2/131},
  \href {https://ui.adsabs.harvard.edu/abs/2013ApJ...765..131K} {765, 131}

\bibitem[\protect\citeauthoryear{{Kopparapu}, {Ramirez}, {SchottelKotte},
  {Kasting}, {Domagal-Goldman}  \& {Eymet}}{{Kopparapu}
  et~al.}{2014}]{Kopparapu2014}
{Kopparapu} R.~K.,  {Ramirez} R.~M.,  {SchottelKotte} J.,  {Kasting} J.~F.,
  {Domagal-Goldman} S.,   {Eymet} V.,  2014, \mn@doi [\apjl]
  {10.1088/2041-8205/787/2/L29}, \href
  {https://ui.adsabs.harvard.edu/abs/2014ApJ...787L..29K} {787, L29}

\bibitem[\protect\citeauthoryear{{Kurashov}, {Ho}  \& {Shen}}{{Kurashov}
  et~al.}{2019}]{Kurashov2019}
{Kurashov} V.,  {Ho} M.,   {Shen} G.,  2019, \mn@doi [Photosynth Res]
  {https://doi.org/10.1007/s11120-019-00616-x}, 141, 151

\bibitem[\protect\citeauthoryear{{Lehmer}, {Catling}, {Parenteau}  \&
  {Hoehler}}{{Lehmer} et~al.}{2018}]{Lehmer2018}
{Lehmer} O.~R.,  {Catling} D.~C.,  {Parenteau} M.~N.,   {Hoehler} T.~M.,  2018,
  \mn@doi [\apj] {10.3847/1538-4357/aac104}, \href
  {https://ui.adsabs.harvard.edu/abs/2018ApJ...859..171L} {859, 171}

\bibitem[\protect\citeauthoryear{Lin, Kuzminov, Park, Lee, Falkowski  \&
  Gorbunov}{Lin et~al.}{2016}]{Lin2016}
Lin H.,  Kuzminov F.~I.,  Park J.,  Lee S.,  Falkowski P.~G.,   Gorbunov M.~Y.,
   2016, \mn@doi [Science] {10.1126/science.aab2213}

\bibitem[\protect\citeauthoryear{{Lingam} \& {Loeb}}{{Lingam} \&
  {Loeb}}{2019}]{Lingam2019a}
{Lingam} M.,  {Loeb} A.,  2019, \mn@doi [\mnras] {10.1093/mnras/stz847}, \href
  {https://ui.adsabs.harvard.edu/abs/2019MNRAS.485.5924L} {485, 5924}

\bibitem[\protect\citeauthoryear{{McKay}}{{McKay}}{2014}]{McKay2014}
{McKay} C.~P.,  2014, \mn@doi [Proceedings of the National Academy of Science]
  {10.1073/pnas.1304212111}, \href
  {https://ui.adsabs.harvard.edu/abs/2014PNAS..11112628M} {111, 12628}

\bibitem[\protect\citeauthoryear{{Miyashita}, {Ikemoto}  \&
  {Kurano}}{{Miyashita} et~al.}{1996}]{Miyashita1996Natur}
{Miyashita} H.,  {Ikemoto} H.,   {Kurano} N.,  1996, \mn@doi [\nat]
  {10.1038/383402a0}, \href
  {https://ui.adsabs.harvard.edu/abs/1996Natur.383..402M} {383, 402}

\bibitem[\protect\citeauthoryear{Nakamura \& Takai}{Nakamura \&
  Takai}{2015}]{Nakamura2015}
Nakamura K.,  Takai K.,  2015, Geochemical Constraints on Potential Biomass
  Sustained by Subseafloor Water--Rock Interactions.
Springer Japan, Tokyo, pp 11--30, \mn@doi{10.1007/978-4-431-54865-2_2}, \url
  {https://doi.org/10.1007/978-4-431-54865-2_2}

\bibitem[\protect\citeauthoryear{Petela}{Petela}{1964}]{Petela_1964}
Petela R.,  1964, \mn@doi [Journal of Heat Transfer] {10.1115/1.3687092}, 86,
  187

\bibitem[\protect\citeauthoryear{Petela}{Petela}{2008}]{Petela2008}
Petela R.,  2008, \mn@doi [Solar Energy]
  {https://doi.org/10.1016/j.solener.2007.09.002}, 82, 311

\bibitem[\protect\citeauthoryear{Pettai, Oja, Freiberg  \& Laisk}{Pettai
  et~al.}{2005}]{PETTAI2005}
Pettai H.,  Oja V.,  Freiberg A.,   Laisk A.,  2005, \mn@doi [Biochimica et
  Biophysica Acta (BBA) - Bioenergetics]
  {https://doi.org/10.1016/j.bbabio.2005.05.005}, 1708, 311

\bibitem[\protect\citeauthoryear{Ptasinski}{Ptasinski}{2016}]{Ptasinski2016efficiency}
Ptasinski K.,  2016, Efficiency of Biomass Energy: An Exergy Approach to
  Biofuels, Power, and Biorefineries.
Wiley, \url {https://books.google.it/books?id=zhysBwAAQBAJ}

\bibitem[\protect\citeauthoryear{{Quintana} et~al.,}{{Quintana}
  et~al.}{2014}]{Quintana2014}
{Quintana} E.~V.,  et~al., 2014, \mn@doi [Science] {10.1126/science.1249403},
  \href {https://ui.adsabs.harvard.edu/abs/2014Sci...344..277Q} {344, 277}

\bibitem[\protect\citeauthoryear{Raven, Kübler  \& Beardall}{Raven
  et~al.}{2000}]{raven2000}
Raven J.,  Kübler J.,   Beardall J.,  2000, \mn@doi [Journal of the Marine
  Biological Association of the United Kingdom] {10.1017/S0025315499001526},
  80, 1–25

\bibitem[\protect\citeauthoryear{{Ritchie}, {Larkum}  \& {Ribas}}{{Ritchie}
  et~al.}{2018}]{Ritchie2018I}
{Ritchie} R.~J.,  {Larkum} A. W.~D.,   {Ribas} I.,  2018, \mn@doi
  [International Journal of Astrobiology] {10.1017/S1473550417000167}, \href
  {https://ui.adsabs.harvard.edu/abs/2018IJAsB..17..147R} {17, 147}

\bibitem[\protect\citeauthoryear{{Rodriguez} et~al.,}{{Rodriguez}
  et~al.}{2020}]{Rodriguez2020}
{Rodriguez} J.~E.,  et~al., 2020, \mn@doi [\aj] {10.3847/1538-3881/aba4b3},
  \href {https://ui.adsabs.harvard.edu/abs/2020AJ....160..117R} {160, 117}

\bibitem[\protect\citeauthoryear{{Rueda}}{{Rueda}}{1973}]{Rueda1973}
{Rueda} A.,  1973, \mn@doi [Space Life Sciences] {10.1007/BF00930358}, \href
  {https://ui.adsabs.harvard.edu/abs/1973SLSci...4..469R} {4, 469}

\bibitem[\protect\citeauthoryear{{Scharf}}{{Scharf}}{2019}]{Scharf2019}
{Scharf} C.,  2019, \mn@doi [\apj] {10.3847/1538-4357/ab12ec}, \href
  {https://ui.adsabs.harvard.edu/abs/2019ApJ...876...16S} {876, 16}

\bibitem[\protect\citeauthoryear{{Schwieterman}, {Reinhard}, {Olson}, {Harman}
  \& {Lyons}}{{Schwieterman} et~al.}{2019}]{Schwieterman_2019ApJ...878...19S}
{Schwieterman} E.~W.,  {Reinhard} C.~T.,  {Olson} S.~L.,  {Harman} C.~E.,
  {Lyons} T.~W.,  2019, \mn@doi [\apj] {10.3847/1538-4357/ab1d52}, \href
  {https://ui.adsabs.harvard.edu/abs/2019ApJ...878...19S} {878, 19}

\bibitem[\protect\citeauthoryear{{Seager} \& {Deming}}{{Seager} \&
  {Deming}}{2010}]{Seager2010ARA&A..48..631S}
{Seager} S.,  {Deming} D.,  2010, \mn@doi [\araa]
  {10.1146/annurev-astro-081309-130837}, \href
  {https://ui.adsabs.harvard.edu/abs/2010ARA&A..48..631S} {48, 631}

\bibitem[\protect\citeauthoryear{Silva, Seider  \& Lior}{Silva
  et~al.}{2015}]{Silva2015}
Silva C.~S.,  Seider W.~D.,   Lior N.,  2015, \mn@doi [Chemical Engineering
  Science] {https://doi.org/10.1016/j.ces.2015.02.011}, 130, 151

\bibitem[\protect\citeauthoryear{{Takizawa}, {Minagawa}, {Tamura}, {Kusakabe}
  \& {Narita}}{{Takizawa} et~al.}{2017}]{Takizawa2017Nat}
{Takizawa} K.,  {Minagawa} J.,  {Tamura} M.,  {Kusakabe} N.,   {Narita} N.,
  2017, \mn@doi [Scientific Reports] {10.1038/s41598-017-07948-5}, \href
  {https://ui.adsabs.harvard.edu/abs/2017NatSR...7.7561T} {7, 7561}

\bibitem[\protect\citeauthoryear{{Tomitani}, {Knoll}, {Cavanaugh}  \&
  {Ohno}}{{Tomitani} et~al.}{2006}]{Tomitani_2006}
{Tomitani} A.,  {Knoll} A.~H.,  {Cavanaugh} C.~M.,   {Ohno} T.,  2006, \mn@doi
  [Proceedings of the National Academy of Science] {10.1073/pnas.0600999103},
  \href {https://ui.adsabs.harvard.edu/abs/2006PNAS..103.5442T} {103, 5442}

\bibitem[\protect\citeauthoryear{{Torres} et~al.,}{{Torres}
  et~al.}{2015}]{Torres2015}
{Torres} G.,  et~al., 2015, \mn@doi [\apj] {10.1088/0004-637X/800/2/99}, \href
  {https://ui.adsabs.harvard.edu/abs/2015ApJ...800...99T} {800, 99}

\bibitem[\protect\citeauthoryear{{Torres} et~al.,}{{Torres}
  et~al.}{2017}]{Torres2017}
{Torres} G.,  et~al., 2017, \mn@doi [\aj] {10.3847/1538-3881/aa984b}, \href
  {https://ui.adsabs.harvard.edu/abs/2017AJ....154..264T} {154, 264}

\bibitem[\protect\citeauthoryear{{Tsapras}}{{Tsapras}}{2018}]{Tsapras2018Geosc}
{Tsapras} Y.,  2018, \mn@doi [Geosciences] {10.3390/geosciences8100365}, \href
  {https://ui.adsabs.harvard.edu/abs/2018Geosc...8..365T} {8, 365}

\bibitem[\protect\citeauthoryear{{Vida}, {Ol{\'a}h}, {K{\H{o}}v{\'a}ri}, {van
  Driel-Gesztelyi}, {Mo{\'o}r}  \& {P{\'a}l}}{{Vida} et~al.}{2019}]{Vida2019}
{Vida} K.,  {Ol{\'a}h} K.,  {K{\H{o}}v{\'a}ri} Z.,  {van Driel-Gesztelyi} L.,
  {Mo{\'o}r} A.,   {P{\'a}l} A.,  2019, \mn@doi [\apj]
  {10.3847/1538-4357/ab41f5}, \href
  {https://ui.adsabs.harvard.edu/abs/2019ApJ...884..160V} {884, 160}

\bibitem[\protect\citeauthoryear{{Villanueva} et~al.,}{{Villanueva}
  et~al.}{2015}]{Villanueva2015S}
{Villanueva} G.~L.,  et~al., 2015, \mn@doi [Science] {10.1126/science.aaa3630},
  \href {https://ui.adsabs.harvard.edu/abs/2015Sci...348..218V} {348, 218}

\bibitem[\protect\citeauthoryear{{Villanueva}, {Smith}, {Protopapa}, {Faggi}
  \& {Mandell}}{{Villanueva} et~al.}{2018}]{Villanueva2018}
{Villanueva} G.~L.,  {Smith} M.~D.,  {Protopapa} S.,  {Faggi} S.,   {Mandell}
  A.~M.,  2018, \mn@doi [\jqsrt] {10.1016/j.jqsrt.2018.05.023}, \href
  {https://ui.adsabs.harvard.edu/abs/2018JQSRT.217...86V} {217, 86}

\bibitem[\protect\citeauthoryear{Wald}{Wald}{1959}]{Wald_1959}
Wald G.,  1959, Sci.Am., 201, 92

\bibitem[\protect\citeauthoryear{{Wolstencroft} \& {Raven}}{{Wolstencroft} \&
  {Raven}}{2002}]{Wolstencroft2002}
{Wolstencroft} R.~D.,  {Raven} J.~A.,  2002, \mn@doi [\icarus]
  {10.1006/icar.2002.6854}, \href
  {https://ui.adsabs.harvard.edu/abs/2002Icar..157..535W} {157, 535}

\bibitem[\protect\citeauthoryear{{Wright}}{{Wright}}{2019}]{Wright2019BAAS}
{Wright} J.,  2019, \baas, \href
  {https://ui.adsabs.harvard.edu/abs/2019BAAS...51c.389W} {51, 389}

\bibitem[\protect\citeauthoryear{Zhang, Li, Yin, Li, Jia, Chen  \& Qiu}{Zhang
  et~al.}{2019}]{Zhang2019_Chl_d}
Zhang Z.-C.,  Li Z.-K.,  Yin Y.-C.,  Li Y.,  Jia Y.,  Chen M.,   Qiu B.-S.,
  2019, \mn@doi [Environmental Microbiology]
  {https://doi.org/10.1111/1462-2920.1 4582}, 21, 1497

\makeatother
\end{thebibliography}
